\documentclass[referee]{raa}
\usepackage{graphicx,times}
\usepackage{natbib}
\usepackage{amssymb,amsmath}
\bibpunct{(}{)}{;}{a}{}{,}
 
\usepackage{hyperref}
\hypersetup{a4paper=true,dvipdfm=true,pagebackref=true, pdftitle = The title of my PDF, pdfauthor = My name, pdfsubject= The subject, pdfkeywords = keyword1 keyword2 keyword3}
\hypersetup{colorlinks = true, linkcolor = green, anchorcolor = red, citecolor = blue, filecolor = red, pagecolor = red, urlcolor = red}
\usepackage{geometry}
\begin{document}

   \title{A Solar Radio Dynamic Spectrograph with Flexible Temporal-spectral Resolution
}

 \volnopage{ {\bf 2017} Vol.\ {\bf X} No. {\bf XX}, 000--000}
   \setcounter{page}{1}

   \author{Qing-Fu Du\inst{1,2}, Lei Chen\inst{1}, Yue-Chang Zhao\inst{1}, Xin Li\inst{1},  Yan Zhou\inst{1}, Jun-Rui Zhang\inst{1}, Fa-Bao Yan\inst{1,2}, Shi-Wei Feng\inst{2}, Chuan-Yang Li\inst{2}, Yao Chen\inst{2}
   }

   \institute{ School of Mechanical, Electrical \& Information Engineering, Shandong University, Weihai, Shandong, Weihai, 264209,
China; {\it dqf@sdu.edu.cn}\\
        \and
             Institute of Space Sciences, Shandong University, Weihai, Shandong, Weihai, 264209, China; {\it hjc-8555@sdu.edu.cn}\\
\vs \no
   {\small Received 2016 August 11; accepted 2017 June 12}
}

\abstract{The observation and research of the solar radio emission have unique scientific values in solar and space physics and related space weather forecasting applications, since the observed spectral structures may carry important information about energetic electrons and underlying physical mechanisms. In this study, we present the design of a novel dynamic spectrograph that is installed at the Chashan solar radio station operated by Laboratory for Radio Technologies, Institute of Space Sciences at Shandong University. The spectrograph is characterized by the real-time storage of digitized radio intensity data in the time domain and its capability to perform off-line spectral analysis of the radio spectra. The analog signals received via antennas and amplified with a low-noise amplifier are converted into digital data at a speed reaching up to 32~k data points per millisecond. The digital data are then saved into a high-speed electronic disk for further off-line spectral analysis. Using different word length (1 k - 32 k) and time cadence (5 ms - 10 s) for the off-line fast Fourier transform analysis, we can obtain the dynamic spectrum of a radio burst with different (user-defined) temporal (5 ms - 10 s) and spectral (3 kHz $\sim$ 320 kHz) resolution. This brings a great flexibility and convenience to data analysis of solar radio bursts, especially when some specific fine spectral structures are under study.
}

   \authorrunning{Q.-F. Du et al. }            
   \titlerunning{A Solar Radio Dynamic Spectrograph with Flexible Temporal-spectral Resolution}  
   \maketitle

%
\section{Introduction}           
\label{sect:intro}

There exist different types of solar radio bursts (defined as a transient enhancement of intensities at radio frequencies), spanning over several decades from the microwave regime to metric-kilometric wavelengths.These radio bursts are released by energetic electrons accelerated during solar eruptions (e.g. flares and coronal mass ejections), through emitting mechanisms like the plasma emission, the thermal bremsstrahlung (free-free) emission, synchrotron emission, as well as the cyclotron maser emission, among other possibilities (\citealt{1950AuSRA...3..387W, 1980panp.book.....M, 1985srph.book.....M}). Of particular interest here is the solar metric radio bursts that are associated with energetic electrons in the corona (see \citealt{2013ApJ...767...29F, 2015SoPh..290.1195F, 2016ApJ...827L...9F};~\citealt{2012ApJ...750..158K, 2016ApJ...830...37K};~\citealt{2014ApJ...787...59C};~\citealt{2016ApJ...830L...2V};~\citealt{2014ApJ...793L..39D, 2015ApJ...812...52D};~\citealt{2016ApJ...826..125K};~\citealt{2016SoPh..291.3369G}, for latest studies). It remains elusive regarding how these electrons are accelerated and how the emission is generated. Observations and scientific studies of solar radio bursts, of interest from its own aspect, can provide valuable diagnostic information on energetic electrons, as well as coronal plasmas and the embedding magnetic field strength and configuration that remain as a major challenge with present techniques. Thus, it is a crucial task of the solar and space physics-space weather communities to develop the state-of-the-art technological system to detect and analyze solar radio bursts.

The bursts are traditionally measured by solar radio spectrographs, which can yield the dynamic spectra within a broad band of frequencies. There exist tens of such systems around the world. Table 1 shows parameters of some solar radio observing systems. Most early solar radio spectrographs provide data with fixed spectral and temporal resolution. One latest radio observational system was built by Yunnan Observatory in 2012, with the maximum temporal resolution reaching up to 2 ms, and the frequency resolution $\sim$ 200 kHz (\citealt{2014NewA...30...68G}). This system employs electronic switches to transfer between the right-handed and left-handed polarized signals, to reduce system cost. Its working frequency is from 70 to 700 MHz with a 11 m parabolic antenna fed by a crossed log-periodic dipole antenna (LPDA).

Many systems listed in Table 1 use the extended Compact Astronomical Low-cost Low-frequency Instrument for Spectroscopy and Transportable Observatory (e-CALLISTO) spectrograph (\citealt{2005SoPh..226..143B, 2009EM&P..104..277B}), which is designed mainly from economic consideration. In addition, the relevant antenna system usually lacks high sensitivity. Therefore, the observational data from CALLISTO systems, although very valuable with its worldwide distribution of 24-hour tracking of the sun, often come with noisy spectral data with temporal and spectral resolutions fixed and relatively low.

Other radio spectrograph systems, e.g., the Green Bank Solar Radio Burst Spectrograph (GBSRBS: http://www.astro.umd.edu/$\sim$white/gb/), the Hiraiso radio spectrograph (\citealt{1994CRLRv..40...85K}), were mostly built ten to twenty years ago, when the electronic devices and computer technology suffered from major limitations. For instance, the analog-to-digital converter (ADC) sampling rate was usually at the rate of tens to a hundred of MHz. According to the Nyquist sampling theorem, the sampling rate should be at least twice the signal bandwidth. Therefore, to observe the wide-band (e.g., 150 $\sim$ 500 MHz) solar radio burst, one has to split the total bandwidth into several segments and assemble them later (e.g. \citealt{2004SoPh..222..167F}). This brings considerable complexity (and thus the noise and cost) to the system and puts a strong limitation on the data quality.


\begin{table}
\caption{Parameters of solar radio observing system in the world}
\begin{center}
\begin{tabular}{|p{2cm}|p{2cm}|p{3.5cm}|p{2cm}|p{2cm}|p{3.5cm}|}
  \hline
  Station name/Country /Construction time & Latitude and longitude/Observation time & Antennas & Observation frequency band (MHz) & Time resolution (s) & Frequency resolution (channel number) (Hz) \\\hline
  AMATERAS /Japan/2010 & E141N38 20-06 UT & $2 * 16.5 * 31$ m of the rectangular paraboloid : $1023 m^{2}$ & 150-500
   & 0.01 & 61 kHz (16384 channels) \\\hline
   BIRS$^{1}$/Australia /1997 & E147S43 20-06 UT & 23 element log-periodic structure & 5-65 & 3 & 268 kHz \\\hline
  Artemis$^{2}$ IV/Greece /1996 & E22N38.5 (06-16UT)	
  & 7 m paraboloid + LPDA (100-650MHz); Anti-V-shaped dipole antenna (20-100MHz) & 20-100, 100-650
& 0.1, 0.01 (270-450) & 1. 630 channels(20-650) 2. 128 channels(270-450)\\\hline
OOTY$^{3}$-Callisto/ India & E76N11 02-12UT & Linear polarization single group LPDA & 45-870 & 0.25 & 200 channels \\\hline
Bleie$^{4}$/ Switzerland & E8N47 (06-16) & 7 m paraboloid, LPDA & 170 - 870 & 0.25 & 200 channels \\\hline
GBSRBS/ US /2004 & W79N38 & 10-80: LPDA £»80-850: 13.7 m  800-3000: 3 m & 10-3000 & 1 &  ~\\\hline
San Vito/ Italy & E18N41 (06-16) & Non-tracking Semi-Bicone, LPDA & 25-75, 75-180 & 3 &	HP8591ESpectrum analyzer \\\hline
Yunnan Observatory$^{5}$ & 102E,24N & 11 m parabolic antenna & 70-700 & 0.002 & 200 kHz \\\hline
Chashan Solar Radio Observatory & 122E,36N & 6m parabolic antenna & 150-500 & 0.005-0.01 & 3 k-320 k (flexible) \\
  \hline
\end{tabular}
\end{center}
1.	Bill Erickson and Hilary Cane's BIRS system on Bruny Island suffered a serious hardware failure in 2015 January and may not be reparable, so there is no new BIRS data since then.\\
2.	http://artemis-iv.phys.uoa.gr/Artemis4\_list.html\\
3.	http://www.e-callisto.org/\\
4.	http://soleil.i4ds.ch/solarradio/\\
5.	http://secchirh.obspm.fr/instruments.php\#ynao\\

\end{table}

Most, if not all, of earlier solar radio spectrographs are providing data with fixed spectral and temporal resolutions. In principle, the temporal and spectral resolutions of these data can still be changed by averaging the data within a certain period (or frequency interval). This always reduces the resolutions of the data. Here we propose a novel design of the solar radio spectrograph, characterized by high-speed real-time storage of the digitized data in the time domain and offline analysis with flexible temporal and spectral resolutions. The system is based on the high-speed wide-band ADCs and hardware computation using FPGA. The sampling speed of the ADC is 1 GHz (12 bits). This allows us to sample the radio signals below 500 MHz with only one ADC, and greatly simplifies the design of the system. It is suggested that this novel system will be helpful to further explore the scientific potential of the radio spectral data, especially for the studies on fine spectral structures.

In the following, we first present two related concepts used in the study, then we introduce the Chashan Solar Radio Observatory (CSO), the data acquisition system (DAS) of the spectrograph, and the calibration methods. The design of the spectrograph is demonstrated by showing the spectral analyses of two solar radio burst events. The last section summarizes this paper.

\section{The spectrograph noise floor and the FFT analysis with different word length}

In this section,we briefly introduce the concepts of the noise floor of a spectrograph and the FFT analysis with different word length, which are closely relevant to our study.

A radio spectrograph can only detect signals with energies above its system noise floor $(p_{n})$, given by\\
\begin{equation}
p_{n}=\lg(kTB)+N_{F}=10\lg(kT)+l0\lg(B)+N_{F}= -174+l0\lg(B)+N_{F}
\end{equation}
where $N_{F}$ is the noise figure of the system, the temperature $T$ is taken to be 290 K, and $B$ is the observing bandwidth, $k$ is the Boltzmann constant $(1.38\times10^{-23}J/K)$, $p_{n}$ is in unit of dBm/Hz. We can see that at a fixed temperature, $p_{n}$ is smaller when $N_{F}$  and $B$ are smaller. It is in general important to reduce the value of $p_{n}$ in order to observe weaker signals.

FFT is a fast algorithm to implement the transform from time domain to frequency domain. The frequency (or spectral) resolution can be determined with the following expression
\begin{equation}
\triangle f = k \times\frac{F_{s}}{N}
\end{equation}
where $F_{s}$ is the total bandwith of the signal, $N$ is the word length of the FFT operation (or the sampling rate, the number of sampling points), and $k$ is the coefficient given by a window function. We see
that $\triangle f$ is inversely proportional to $N$.
 To detect a signal with specific fine spectral structures, high-enough spectral (and temporal) resolution is required. Yet, it should be noted that both the temporal and spectral resolutions can not be increased unlimitedly due to the limited energy of the signal. The window function, usually bell-shaped, is to reduce signal distortions and spectral energy dispersions, which may be introduced between successive frames due to the FFT operations with limited data. In addition, with a larger $N$ and thus a smaller $\triangle f$, the power of the background emission (i.e., the noise) shall decrease accordingly, this effectively reduces the noise floor, as illustrated in the spectral analysis.

\section{The CSO, DAS, and CALIBRATION methods}

\subsection{A brief introduction to the CSO}

The construction of the Chashan solar radio observatory (CSO) started from 2015. The first light (observing a solar radio burst) was received in July 2016. The observatory is located at the Chashan mountain of Rongcheng City, the southernmost part of the Jiaodong Peninsula, and operated by the Laboratory for Radio Technologies, Institute of Space Sciences, Shandong University. The spectral band of interest here is from 150 to 500 MHz, observed with a 6 m parabola fed by a crossed LPDA, which can track the Sun automatically from 6 AM to 5 PM local time in summer and from 8 AM to 5 PM in winter. The northern part of the site is the mountain with an altitude up to 500 m; the southern part is the Yellow Sea; the eastern and western parts are partially surrounded by mountains. Therefore, this site has the advantage of low radio interference (RFI). It is a factor extremely important for a successful observation of solar radio bursts at the metric wavelength, as testified by the on-going observations.

\subsection{Data Acquisition System}

The core part of the data acquisition system (DAS) is the high-speed acquisition card, with a 2-channel synchronous 12 bit 1 GSPS (Gigabit Samples Per Second) ADC and 4 GB of high-speed DDR3 cache memory, and a Xilinx XC6VSX315T FPGA chip. The card supports 8-channel and up to 32k ($N$) synchronous real-time FFT processing. The system also supports the PCIE2.0 x8 bus with data transmission speed as large as 2.5 GB/s. The programs for the purpose of this study are executed on this card. The following diagram shows the main functions of the spectrograph (see the large squared area).

\begin{figure}[htbp]
   \centering
   \includegraphics[width=14.0cm, angle=0]{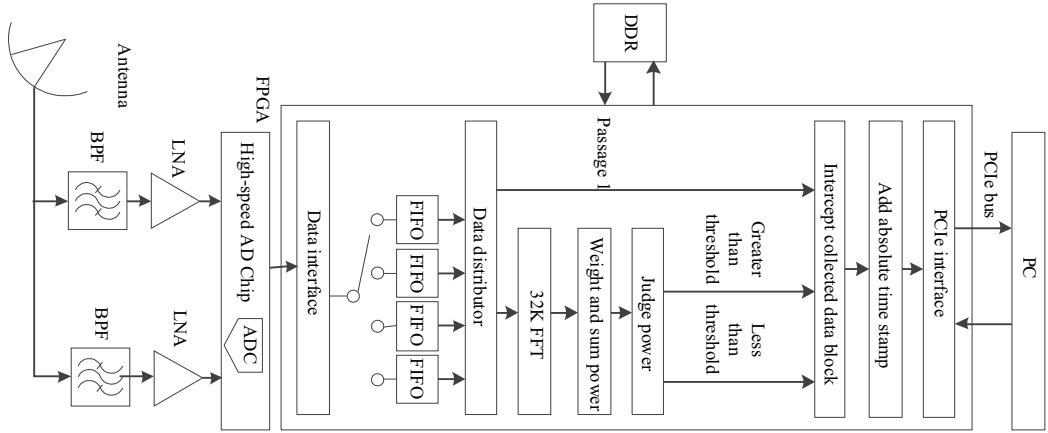}
   \caption{The system block diagram. }
   \label{Fig1}
\end{figure}

The signals, received by the antenna and processed by the low-noise amplifier (LNA), are converted into digital data via the ADC. We use 4 FIFO (First-In First-Out) registers to decelerate the data flow to $\frac{1}{4}$ of the original speed. The FFT functions for the data received through the 4 registers are performed independently. The obtained results are then combined to form a full spectrum. To reduce the influence of RFI, we find out the known channels of RFI and set the signal zero there. Then we sum up the total energy within the observing band to give the weighted total power of the spectrum ($It$).When the obtained $It$ is larger (smaller) than a prescribed threshold, the system works at the burst (normal) mode. The threshold is determined by observational tests.

With the normal mode, the cadence of data collection is relatively low. For instance, we can set to collect 32 k data points per second. While with the burst mode, the system works to its full potential, and 32 k data points will be collected per millisecond. In each mode, the real-time dynamic spectra will be saved (together with the time stamps) to produce a synoptic spectrum. In the meantime, the output digital data (in the time domain) of the ADC (together with the time stamps) will be saved onto a high-speed electronic hard disk for further off-line analysis. Note that in the above setup of the burst mode, one can obtain a huge amount of data with $\sim 96$ GB in 10 minutes.

To store the huge amount of the digitized time-domain data during the burst mode, we set up a 4 GB DDR3 cache memory with two data blocks (A and B). The volume of A or B is configured according to the amount of the data that need to be saved. The FPGA chip is linked directly to the high-speed PCIE bus of the computer. Data from the ADC and the FFT results of the FPGA are first buffered in the blocks, and then uploaded to the computer via the ping-pong method. In other words, the data are first saved in the block A, and uploaded when A is full; during the uploading process, data will be saved in block B, and uploaded when it is full; then data will be saved in A again, and so on. In addition, we need to make sure that the uploading speed is always larger than the data acquisition speed to avoid loss of data. The data package on the computer will be named according to the data and the exact time of the package. Usually, the system is set to collect and save the data automatically during daytime, and after the sunset, the data will be examined manually to remove the data of no interest (e.g. without any signature of radio bursts) to save space.

The control software of the spectrograph is developed with the Microsoft Foundation Class Library and installed on the host computer. Through the software, the user can select data channels, set the temporal resolution, the power threshold to determine the running mode of the system (i.e., normal or burst), and file operations, etc.

\subsection{Data Calibration}

We currently designed two calibration methods. The first one has been implemented and the second one is still under construction. The former is presented as follows.

The Learmonth solar radio observatory releases calibrated solar radio flux intensity at several discrete frequencies (from 245 MHz to 15400 MHz). One can use these open-access data for calibration. The 245 MHz is within the observational range of CSO spectrograph, therefore the data at this frequency can be calibrated. To do this, we apply the following algorithm to the value $R$ that is the spectral intensity at the selected frequency in arbitrary unit,
\begin{equation}
F_{A} = A\times \lg[(R-B)\times C]
\end{equation}
by adjusting the parameters $A$, $B$, and $C$, we make sure that the deduced $F_{A}$ within a certain period is basically the same as that given by Learmonth ($F$). The comparison of the two curves can be seen from Figure 2. We have $A = 1.0040$, $C = 0.2890$ and $B= 572.5720$. To test this calibration method, we repeat the above calculation (with the same set of coefficients) to the radio flux data obtained the next day (July 19, 2016). A consistent result is found (see Figure 2b). This indicates the success of this calibration method. In practice, the coefficients ($A$, $B$, $C$) should be corrected every day to make sure a proper calibration.
\begin{figure}[htbp]
   \centering
   \includegraphics[width=14.0cm, angle=0]{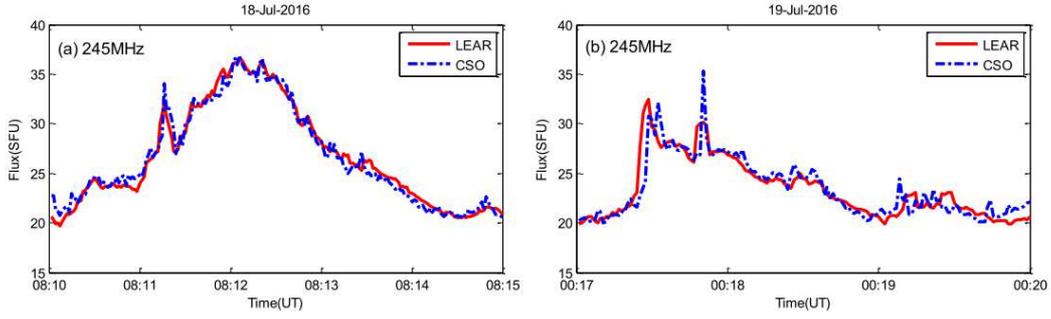}
   \caption{Comparison of the calibrated data obtained by CSO spectrograph and those released by the Learmonth solar radio observatory in Australia at 245 MHz.}
   \label{Fig2}
\end{figure}

Another calibration method is usually referred to as the Y-factor method (see, e.g., \citealt{2015ApJ...808...61T}). This method utilizes a standard noise source with a known power of output signal in white noise. With this method, the front electronic system and the receiver (all electronic devices after the antenna) can be calibrated (see Figure 3). The antenna properties should be tested and calibrated independently. A noise generator with a 30 dB ENR (Excess Noise Ratio) and a 50 Ohm resistor are used for the calibration.
\begin{figure}[htbp]
   \centering
   \includegraphics[width=14.0cm, angle=0]{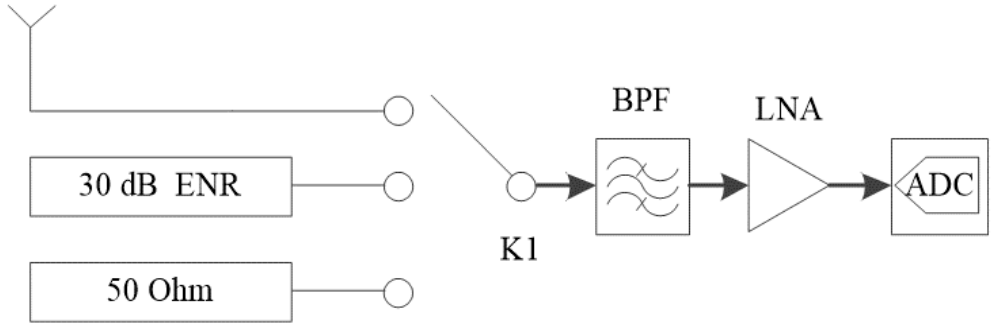}
   \caption{The Y-factormethod of the spectrographcalibration, using a standard noise source with 30 dB ENR.}
   \label{Fig3}
\end{figure}

The temperature $(T_{\textrm{ENR}})$ of the 30 dB ENR noise source can be calculated through\\
\begin{equation}
P_{\textrm{ENR}} = 10\lg(\frac{T_{\textrm{ENR}}}{T_{0}}-1)
\end{equation}
where $T_{0}$ is the room temperature and $P_{\textrm{ENR}}=30$ dB. The receiver reads $P_{T_{0}}$ when connected to the 50 ohm resistor, and reads $P_{\textrm{ENR}}$ when connected to the noise source, and the system temperature of the receiver is $T_{\textrm{SYS}}$. One has
\begin{equation}
P_{\textrm{ENR}}  = \frac{T_{\textrm{ENR}}-Y\times T_{0}}{Y-1}
\end{equation}
\begin{equation}
G = \frac{T_{\textrm{ENR}}}{kB(T_{\textrm{SYS}}-T_{0})}
\end{equation}
where $G$ is the gain of the receiver and $Y=\frac{P_{T_{\textrm{ENR}}}}{P_{T_{0}} }$. Note that this calibration method should be used together with the test and calibration of the antenna-feed system (not finished yet).

\section{OFFLINE ANALYSIS OF THE DATA}

As introduced earlier, the spectral resolution is directly determined by the word length $N$ of the FFT analysis. When $N$ increases by a factor of 10, so does the spectral resolution. For example, with a sampling frequency being 1 GSPS, $N = 32$ k corresponds to a frequency resolution of $\sim 30$ kHz, and $N = 320$ k corresponds to a frequency resolution of $\sim 3$ kHz. In addition, with a larger $N$, smaller values of the system noise floor can be obtained. In the following we demonstrate this effect of $N$ on the FFT analysis with two solar radio burst events recorded by CSO spectrograph on July 18, 2016 and July 19, 2016. The data sampling is at a cadence of 10 ms with 32 k data points being collected at each sampling.

Figure 4 shows the comparison between the dynamic spectra with $N=1$ k and $32$ k
($\triangle f= \sim 1$ MHz and $\sim 30$ kHz) with the same temporal resolution ($\triangle t = 250$ ms). The full spectra from 150 to 500 MHz are shown in upper panels with different contrast. Note the color of the background emission of the two panels has been adjusted to be close to each other, for the convenience of comparison. The lower two panels of Figure 4 show the spectra observed from 225 to 235 MHz. In both panel (b) and (d), the background level is weaker by about 15 dBm due to a larger $N$ (32 k versus 1 k). This figure clearly presents the effect of $N$ on the noise floor of the system.

To show the comparison in a more quantitative way, we plot the flux curves at 305 MHz in the upper panels of Figure 5. The data are shown for different temporal resolutions (50 ms and 250 ms) as well as different FFT word length (1 k and 32 k). We see that the data with 250 ms present a slightly lower noise floor by $\sim 2$ dBm than that of $\triangle t = 50$ ms (with the same $N$), the data with $N=1$ k present a much larger noise floor by $\sim 15$ dBm than that of $N=32$ k (with the same $\triangle t$).

In the lower panels of Figure 5, we show the SNR (Signal Noise Ratio)of the data. The SNR is calculated with the following equation,
\begin{equation}
SNR = 10\lg(\frac{P_{s}}{P_{n}})(dB)
\end{equation}
where $P_{s}$ is the power of the signal and the noise floor of the spectrograph $P_{n}$ is represented by the root mean square (RMS) of the fluctuating power of the background signal. We see that although the noise floor of different parameters is very different, the SNRs are quite similar to each other. This is due to the fact that the powers of both the signals and the background noises change accordingly with the spectral and temporal resolutions.

\begin{figure}[htbp]
   \centering
   \includegraphics[width=14.0cm, angle=0]{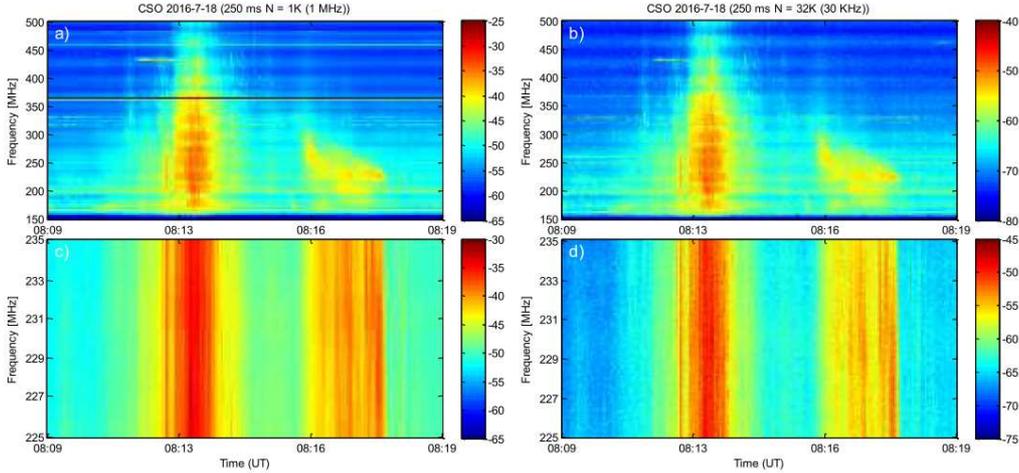}
   \caption{The solar type III and type IV radio bursts on July 18, 2016 observed by the CSO, shown for comparison of spectra with different FFT word length (1 k and 32 k) with the same temporal resolution (250 ms). The spectral resolution is about 1 MHz for $N = 1$ k and about 30 kHz for $N = 32$ k, as written in the parenthesis.}
   \label{Fig4}
\end{figure}

\begin{figure}
   \centering
   \includegraphics[width=14.0cm, angle=0]{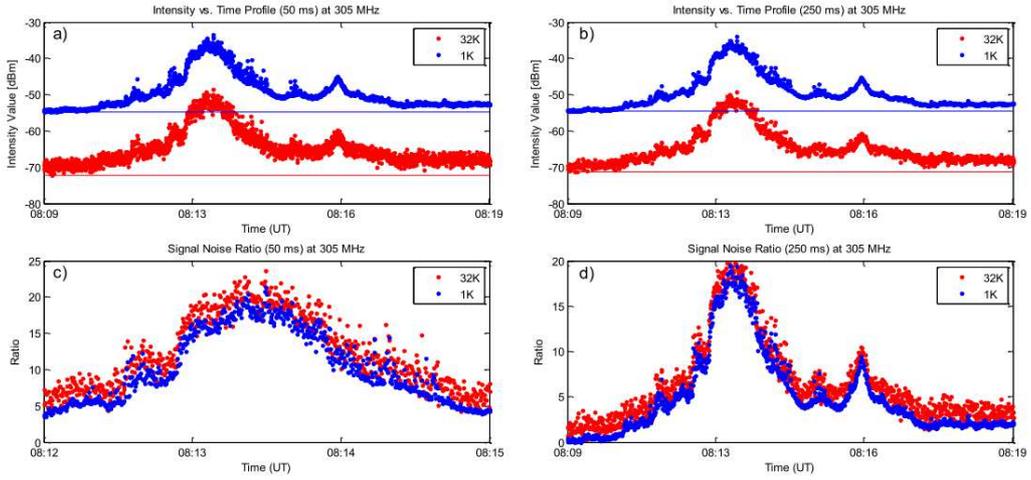}
   \caption{Panel (a) and (b) show the flux intensity curves at 305 MHz, with different temporal resolutions (50 ms and 250 ms) and different FFT word length (1 k and 32 k), and (c) and (d) show the corresponding SNRs. The horizontal lines represent the corresponding noise floor given by the RMS of the background intensity.}
   \label{Fig5}
\end{figure}

To demonstrate the capability of CSO spectrograph in identifying fine spectral structures, in Figure 6 we show two sets of dynamic spectra observed on July 18, 2016 (the same event reported above) and July 19, 2016, respectively. Each event is shown for $N = 1$ k and $N = 32$ k with the spectral resolution being 1 MHz and 30 kHz. The contrast levels of the spectra with different $N$ have been adjusted to reveal similar color of the background emission.

In panels (a) and (d) of Figure 6, we see that there exist observable mosaic patterns due to the low frequency resolution, while in panels (b) and (e) such mosaic patterns are removed due to the increased spectral resolution. In addition, the filamentary fine structures and their drifting trend/rate can only be clearly discernible from panels (b) and (e). See the region given by the white rectangles. Panels (c) and (f) show the dynamic spectra the frequency resolution of 3 kHz ($N = 320$ k) that is the highest one of CSO spectrograph. Note that in order to have enough energy at each data point, we use a low time resoltuion of 960 ms.Compared with Figure4, the noise floor is much lower at the frequency resolution of 3kHz, by 10 dBm than that at $N = 32$ k and $\triangle t= 250$ ms, and by 25 dBm than that at $N = 1$ k and $\triangle t= 250$ ms. Nevertheless, the energy at each data point is quite limited by this super-high spectral resolution. This affects the quality of the spectra, indicating that indeed the resolutions (spectral or temporal) cannot be increased unlimitedly.

\begin{figure}[htbp]
   \centering
   \includegraphics[width=14.0cm, angle=0]{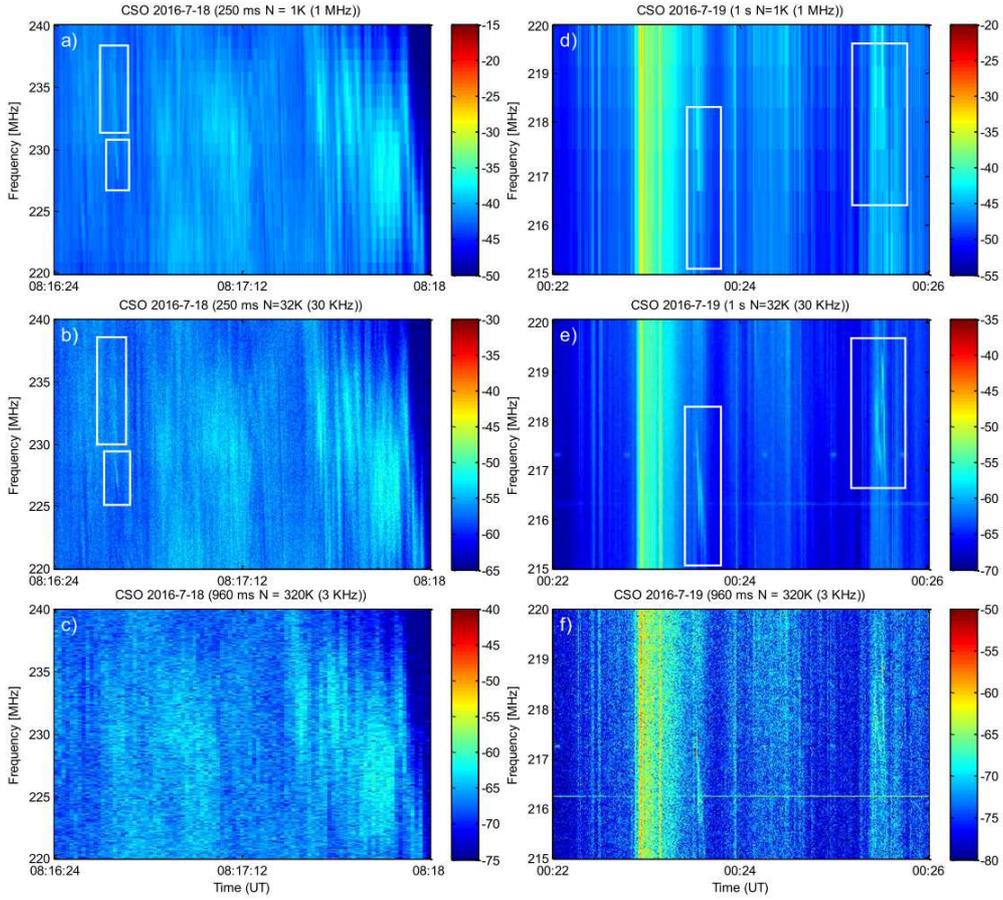}
   \caption{Two events observed on July 18 (a-c), 2016and July 19 (d-f), 2016. The contrast levels of the spectra with different $N$ have been adjusted to reveal similar color of the background emission.The FFT word length $N$ is taken to be 1 k, 32 k, and 320 k for the upper, middle and lower panels, respectively, with different temporal resolutions. The time resolution of the left panels is 250 ms (a, b), and 960 ms (c), and the time resolution of the right panels is 1 s (d, e) and 960 ms (f), respectively. The white rectangle regions present the area of interest.}
   \label{Fig6}
\end{figure}

Figure 7 is a comparison of the dynamic spectra with different temporal resolutions ($\triangle t = 1$ s for panel (a) and $\triangle t = 10$ ms for panel (b)), and the same spectral resolution (1 MHz). We see that the mosaic patterns are present in Figure 7 (a) of a lower temporal resolution while disappears in Figure 7(b) of a higher temporal resolution. In addition, in Figure 7(b) a lot of drifting fine-filamentary spectral structures present.

In summary, with the above comparisons of dynamic spectra with different spectral and temporal resolutions, we show the potential capability of using different combinations of temporal and spectral resolutions in investigating the fine spectral structures of a radio burst. This capability will be especially useful when details of fine spectral structures are under study.

\begin{figure}[htbp]
   \centering
   \includegraphics[width=14.0cm, angle=0]{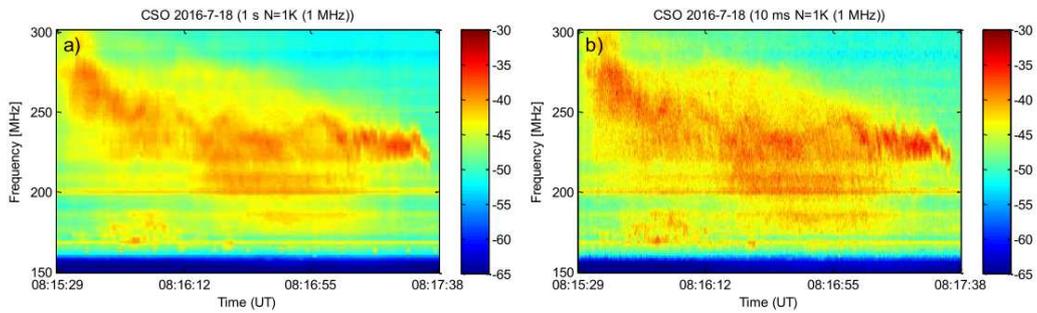}
   \caption{The dynamics spectra with different temporal resolution: (a) ¦¤t = 1 sand (b) ¦¤t = 10 ms, using the same spectral resolution $(\triangle f = 1$ MHz) }
   \label{Fig7}
\end{figure}

\section{CONCLUSION}

There are dozens of solar radio spectrographs around the world that provide continuous data of dynamic spectra of solar radio bursts. These data contain important information about energetic electrons and relevant processes of solar eruptions, particle acceleration through shocks or magnetic reconnection, and electro-magnetic radiations. Some spectrographs were built one to two decades ago with the electronic and digital technologies suffering from serious limitations. In addition, most systems provide data at fixed spectral and temporal resolutions. Here we present a novel design of solar radio spectrograph, characterized by the real-time storage of the radio intensity data in the time domain and further off-line analysis of the dynamic spectra with user-defined flexible temporal and spectral resolutions. With the off-line analysis, the system can achieve a frequency resolution of 3 kHz, that is the highest-ever resolution, ever reported, to our knowledge. The spectrograph was installed at the CSO station, which has the advantage of very low radio interference. This allows the system to work properly and record clean details of solar radio bursts at the metric wavelength.

The system runs in two modes, the normal mode and the burst mode, determined automatically with a prescribed threshold of the total intensity. The merit of the design is demonstrated with a detailed analysis of two radio burst events recorded by CSO spectrograph. Calibration methods of the system are also briefly introduced. The design of the spectrograph brings a great flexibility to the analysis of the solar radio bursts. It is particularly helpful when specific fine spectral structures are of major interest, which can be easily missed from detection if the spectral-temporal resolutions were fixed as usually done with other spectrographs.

\normalem
\begin{acknowledgements}
This work was supported by the National Natural Science Foundation of China (41331068, 11503014, and U1431103) and the China Postdoctoral Science Foundation (2016M600538).
\end{acknowledgements}

\bibliographystyle{raa}
\bibliography{bibtex}

\end{document}